\documentclass[compsoc,conference,a4paper,10pt,times]{IEEEtran}
\IEEEoverridecommandlockouts
\usepackage{cite}
\usepackage{amsmath,amssymb,amsfonts}
\usepackage{algorithmic}
\usepackage{graphicx}
\usepackage{textcomp}
\usepackage{bmpsize}
\usepackage{xcolor}
\usepackage{lipsum}
\usepackage{booktabs}
\usepackage{graphicx}
\usepackage{tcolorbox}
\usepackage{xurl}
\usepackage[T1]{fontenc}
\usepackage[utf8]{inputenc}
\usepackage[colorlinks=true,urlcolor=black]{hyperref}
\def\BibTeX{{\rm B\kern-.05em{\sc i\kern-.025em b}\kern-.08em
    T\kern-.1667em\lower.7ex\hbox{E}\kern-.125emX}}

\makeatletter 
\newcommand{\linebreakand}{%
  \end{@IEEEauthorhalign}
  \hfill\mbox{}\par
  \mbox{}\hfill\begin{@IEEEauthorhalign}
}
\makeatother 

\begin{document}

\title{LLM in the Shell: Generative Honeypots\\ }

\author{\IEEEauthorblockN{1\textsuperscript{st} Muris Sladić}
\IEEEauthorblockA{\textit{Czech Technical University in Prague}\\
Prague, Czech Republic \\
sladimur@fel.cvut.cz}
\and
\IEEEauthorblockN{2\textsuperscript{nd} Veronica Valeros}
\IEEEauthorblockA{\textit{Czech Technical University in Prague}\\
Prague, Czech Republic \\
veronica.valeros@fel.cvut.cz}
\and
\IEEEauthorblockN{3\textsuperscript{rd} Carlos Catania}
\IEEEauthorblockA{\textit{CONICET, UNCuyo}\\
Mendoza, Argentina \\
harpo@ingenieria.uncuyo.edu.ar}
\and

\linebreakand

\IEEEauthorblockN{4\textsuperscript{th} Sebastian Garcia}
\IEEEauthorblockA{\textit{Czech Technical University in Prague}\\
Prague, Czech Republic \\
sebastian.garcia@agents.fel.cvut.cz}
}

\maketitle

\begin{abstract}
Honeypots are essential tools in cybersecurity for early detection, threat intelligence gathering, and analysis of attacker's behavior. However, most of them lack the required realism to engage and fool human attackers long-term. Being easy to distinguish honeypots strongly hinders their effectiveness. This can happen because they are too deterministic, lack adaptability, or lack deepness. This work introduces shelLM, a dynamic and realistic software honeypot based on Large Language Models that generates Linux-like shell output. We designed and implemented shelLM using cloud-based LLMs. We evaluated if shelLM can generate output as expected from a real Linux shell. The evaluation was done by asking cybersecurity researchers to use the honeypot and give feedback if each answer from the honeypot was the expected one from a Linux shell. Results indicate that shelLM can create credible and dynamic answers capable of addressing the limitations of current honeypots. ShelLM reached a TNR of 0.90, convincing humans it was consistent with a real Linux shell. The source code and prompts for replicating the experiments have been publicly available.
\end{abstract}

\begin{IEEEkeywords}
Large Language Models, honeypots, shelLM
\end{IEEEkeywords}

\section{Introduction}
Honeypots allow security researchers and defense teams to detect, monitor, and log attacks on their systems and networks~\cite{ilgSurveyContemporaryOpensource2023}. In theory, this detection information could be used for behavioral analysis of attackers, but in reality, it is mostly used for gathering threat intelligence and telemetry~\cite{enisa}. We believe that the main cause for this limited use is that (i) most attacks from the Internet are automated, and (ii) current shell-based honeypots are easy to identify for a human attacker~\cite{softwarehoneypots}. Moreover,  the process of successfully understanding, learning, and recognizing attackers' behaviors takes considerable effort and requires large amounts of data.   

We study the problem of honeypots not being sufficiently good at mimicking system shells (such as Bash) to make human attackers believe long enough that they are interacting with a real system~\cite{1495930}. Current honeypots are usually too limited in possible actions, their file systems too generic, their answers too hard-coded, and the overall system does not look complex enough~\cite{Simple}.

Our work proposes \texttt{shelLM}, a shell-based honeypot software using Large Language Models (LLMs). The aim of shelLM is to generate a credible and realistic Linux shell that is engaging for attackers and, therefore, may delay the realization that they are not interacting with a real Linux shell. LLMs can create all the needed information, from file system structure to on-demand file content. 

\texttt{ShelLM} was evaluated in its capacity to generate output as is expected from a Linux shell. This was done by asking 12 cybersecurity experts to use the system and give feedback. Results show a 0.90 TNR, meaning that 90\% of the time, the answers were assessed to be consistent with a Linux shell. We show that by using good, prompt engineering techniques, we can use LLMs to create realistic honeypot systems.

The key features of \texttt{shelLM} are: (i) the content of a session is always transferred into the new session of the same user to keep future consistency, (ii) the use of the chain-of-thought prompt technique~\cite{wei2023chainofthought}, and (iii) the use of prompts with precise instructions to avoid certain pitfalls.

This research has two main contributions: first, it presents human-tested empirical evidence that supports the use of LLMs for building credible honeypots; second, the publication of an LLM-based honeypot software. 

\section{Related work}

In the field of cyber-defense, honeypots are used as deception tools to attract potential attackers and to keep them occupied and away from other devices while gathering information about their techniques and methods~\cite{yuill2000intrusion}.
Moreover, they are effective tools for detecting insider threats~\cite{1254322}. Based on their level of interaction, honeypots are usually classified as low-interaction, high-interaction~\cite{provos2004virtual}, or medium-interaction~\cite{wicherski2006medium}. Cowrie is a well-known medium-interaction honeypot used by many organizations~\cite{Cowrie}. It is designed to log attacks and shell interactions the attacker performs by emulating a Linux system. However, Cowrie is quite static since it follows a configuration file.

The application of artificial intelligence and Natural Language Processing (NLP) techniques in the context of honeypot generation is an emerging research topic. In particular, the dynamic capabilities of AI techniques seem to be adequate for dealing with the intrinsic variation of attacker behaviors. For instance, the authors of~\cite{6866707} employ a Reinforcement Learning (RL) approach to design a self-learning honeypot. The RL-based honeypot can adapt its strategy by interacting with attackers, providing a more dynamic defense mechanism that learns from each encounter. This paves the way for more adaptive systems to better understand and counter evolving cyber threats. Furthering the concept of adaptability, the study in~\cite{inproceedings} proposes the creation of self-adapting honeypots based on a game-theoretical approach between the attacker and the honeypot. This offers a more complex and interactive defensive landscape, making it difficult for attackers to identify and bypass the honeypot.

More recently, researchers applied NLP techniques to extract information from honeypot logs. The work of~\cite{9799396} employs various NLP techniques to cluster different types of cyber attacks, providing valuable insights into attacker behaviors and tactics. Such information can then be used to fine-tune honeypot configurations and improve the overall cybersecurity posture. A similar approach is presented in~\cite{gpt-2c}, where a fine-tuned LLM (GPT-2C) model parses dynamic logs generated by Cowrie. This fine-tuned model aids in real-time analysis and threat identification, showcasing how machine-learning algorithms can enhance the effectiveness of honeypots.

There are only a few studies on applying NLP models for implementing the content of honeypots. In a preliminary analysis~\cite{mckee2023chatbots}, authors propose an approach for interfacing with ChatGPT, demonstrating how to formulate prompts that can mimic the behavior of Linux shell commands on various operating systems. More recently, a new LLM-based web honeypot software called Galah~\cite{Galah} was introduced, with abilities to dynamically respond to arbitrary HTTP requests.

\section{Methodology and Implementation}


We created and deployed a honeypot software called \texttt{shelLM} using Large Language Models. \texttt{shelLM} is implemented in Python and uses a cloud-based LLM. It is designed to work seamlessly with a typical SSH setup, facilitating user interactions via this connection.

The process involved: (i) designing specific prompts and applying well-known prompt engineering techniques, (ii) implementing the honeypot software using these prompts, (iii) connecting \texttt{shelLM} to an actual SSH server, (iv) designing experiments, and (v) asking security researchers to evaluate \texttt{shelLM}.


\subsection{Prompt Engineering Technique}
To guide the behavior of \texttt{shelLM}, a cloud-based LLM  is given an initial prompt at the beginning of each user's session. This initial prompt contains a \textit{personality prompt} and a \textit{thinking process prompt}. The \textit{personality prompt} instructs the LLM to be (i) precise, (ii) realistic, (iii) not to disclose its internal details, (iv) describes the expected behavior of a Linux shell, and (v) provides examples of desired output under certain situations. The \textit{thinking process prompt} instructs the LLM to \textit{'think step-by-step'} following the Chain of Thought (CoT) prompt technique~\cite{wei2023chainofthought}, in combination with the \textit{few-shot technique}~\cite{brow_few_shot_2020}, and repeated orders to enforce this behavior. 

\subsection{Session Management}
\begin{figure}[h]
    \centering
    \includegraphics[width=0.45\textwidth]{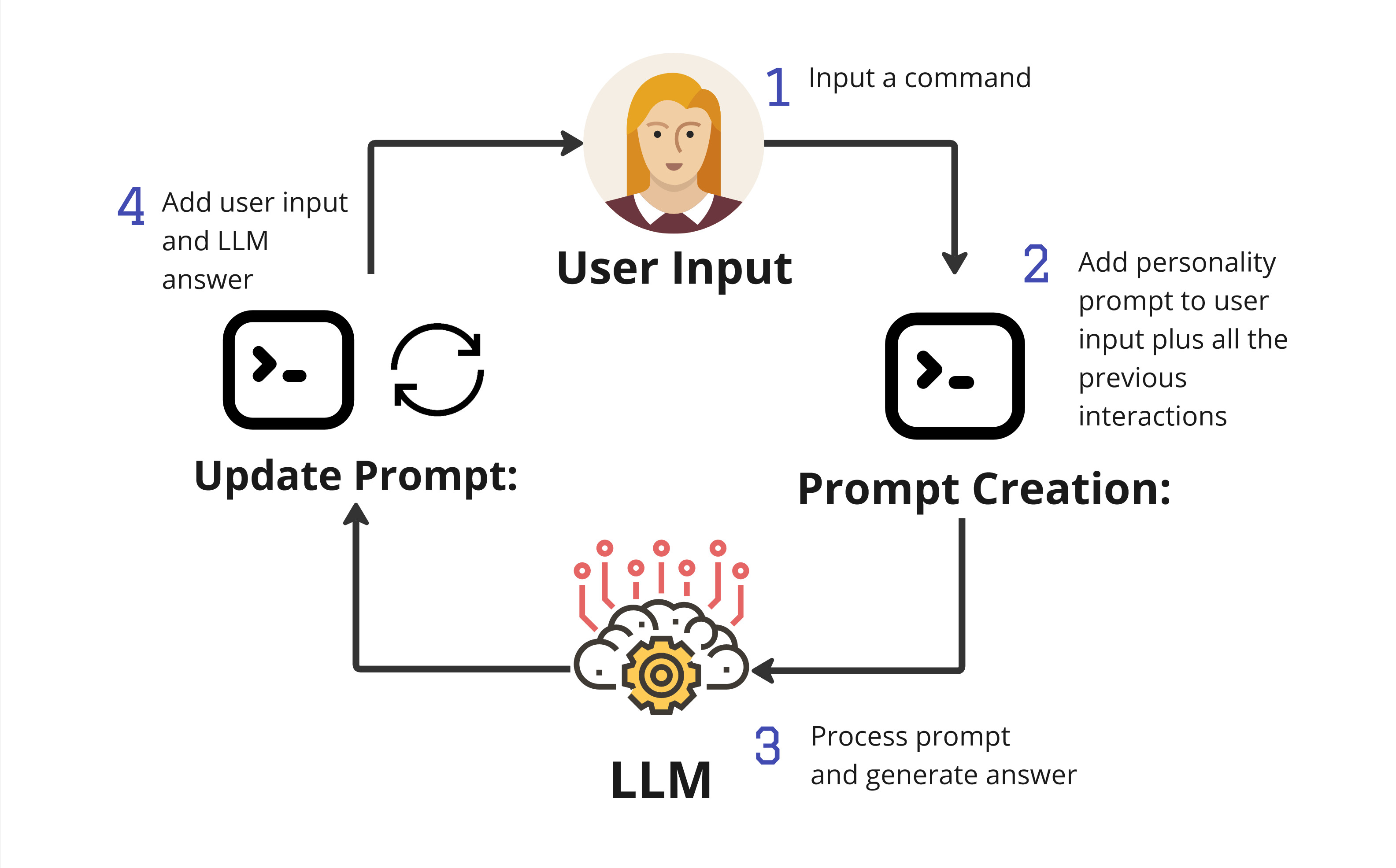}
   \caption{Prompt construction during a shelLM session}
    \label{fig:shellm-session}
\end{figure}

A session in \texttt{shelLM} contains all the inputs and outputs between a user and \texttt{shelLM} but also involves a prompt engineering technique. The operation is similar to the approach used for building chat-bots on top of an LLM, as shown in Figure~\ref{fig:shellm-session}. 

The session begins with the user interacting with the LLM. This is the initialization point where the \textit{personality prompt} is first introduced. Then, a prompt engineering process is repeated for each new input until the session ends. The prompt construction process consists of:

\begin{enumerate}

\item \textbf{User Input:} The user inputs a command.

\item \textbf{Prompt Creation:} This prompt integrates the original personality prompt, the history of all \textit{previous interactions} (both user inputs and responses), and the new user command.

\item \textbf{LLM Processing:} The LLM processes this comprehensive prompt and generates a response.

\item \textbf{Update Prompt:} The user command and the LLM response are then added to the history of \textit{previous interactions}.
\end{enumerate}

\subsection{Model and Consistency Techniques}
The LLM model used was GPT-3.5-turbo-16k from OpenAI~\cite{chatgpt}. The model used a temperature of $0$, a maximum of $800$ tokens per response, and the rest of the parameters left by default. 

To keep consistency between multiple sessions from the same user, the complete content of each session is saved to an external file and used as part of the initial prompt for the next session from the same attacker. Attackers are identified by their IP addresses and login usernames, so upon reconnection, the same attacker can continue the interaction in the same state. This means, for example, that if an attacker creates a file in the honeypot and later wants to access that file, they will be able to do so, even if they reconnect the next day. Due to the non-deterministic nature of LLMs sometimes the same command might not return the same response, which can be beneficial in some cases. For instance, some commands rely on varying responses each time they are executed.  This variability is useful for simulating the conditions observed in real active environments with multiple users. However, for most commands, the model is prompted to generate the same output every time.

The initial prompt and the whole history are saved as part of the context for the LLM. If the context reaches the maximum of 16k tokens, then the history is deleted, restoring the initial personality prompt. In our experiments, the context took a long time to be filled and didn't present an issue for the attackers. However, in future work, we will deal with the history management in a better way. 

\section{Demonstrating shelLM in Action}
\begin{figure}[h]
    \centering
    \includegraphics[scale=0.33]{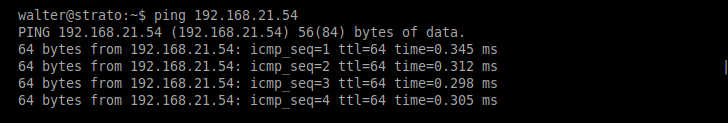}
    \caption{ShelLM output for \texttt{ping} command.}
    \label{fig:shellm_cmd_ping}
\end{figure}

\begin{figure}[h]
    \centering
    \includegraphics[scale=0.33]{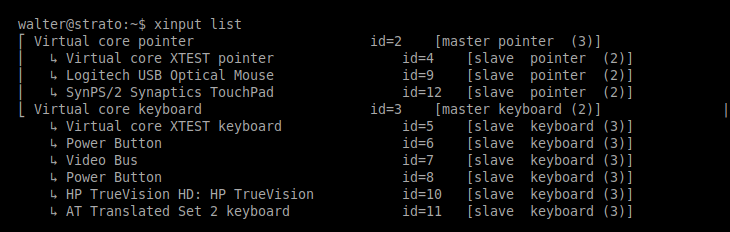}
    \caption{\texttt{shelLM} output for the \texttt{xinput} command.}
    \label{fig:shellm_cmd_xinput}
\end{figure}

\begin{figure}[h]
    \centering
    \includegraphics[scale=0.33]{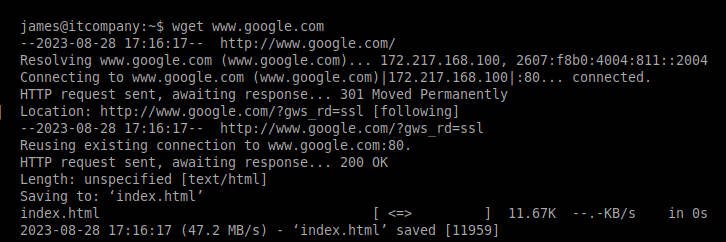}
    \caption{\texttt{shelLM} output of a \texttt{wget} command.}
    \label{fig:shellm_cmd_wget}
\end{figure}

An example output generated by \texttt{shelLM} can be observed in Figure~\ref{fig:shellm_cmd_ping}. The Figure shows the attacker executing the \texttt{ping} command and \texttt{shelLM} automatically generating the output in real-time, showing how \texttt{shelLM} emulates network connectivity. The output of ping is printed line by line to simulate delays. Figure~\ref{fig:shellm_cmd_xinput} shows the output of the \texttt{xinput} command, mimicking input device configurations. Figure~\ref{fig:shellm_cmd_wget} shows the attacker issuing the \texttt{wget} command, illustrating the honeypot capabilities for replicating network interactions with other servers. These examples help to highlight the precision and variability with which the LLM honeypot simulates genuine system behaviors. Furthermore, these commands are not supported by standard honeypots, such as Cowrie, and this makes a major differentiation between them.

\texttt{ShelLM} is published on GitHub at~\url{https://anonymous.4open.science/r/shelLM}. The tool requires a valid OpenAI API key for GPT-3.5-turbo-16k.

\section*{Setup of Evaluation Experiments}
Since \textbf{the goal of this research is to investigate whether LLM-based shell honeypots can generate outputs indistinguishable from those of a real Linux shell, and not to assess participants' ability to detect a honeypot}, the evaluation methodology has to be carefully constructed. Our approach was to tell the evaluation participants that they were connecting to a honeypot but to ask them if the answer from the honeypot looked realistic or not and why. This way a small bias is introduced, but it was possible to understand if and why the output looked convincing.

We contacted 12 participants by email, gave them unique users and passwords, and gave them the following requirement: \textit{Your goal in this experiment is to identify which output of the system helps you detect it as a honeypot. Each command is going to give an output and you need to explain how each output helps or does not help you identify the system as a honeypot. For each command please copy it in the email and briefly explain why it did or did not help you realize it is a honeypot.}

\begin{figure}[h]
    \centering
    \includegraphics[width=0.45\textwidth]{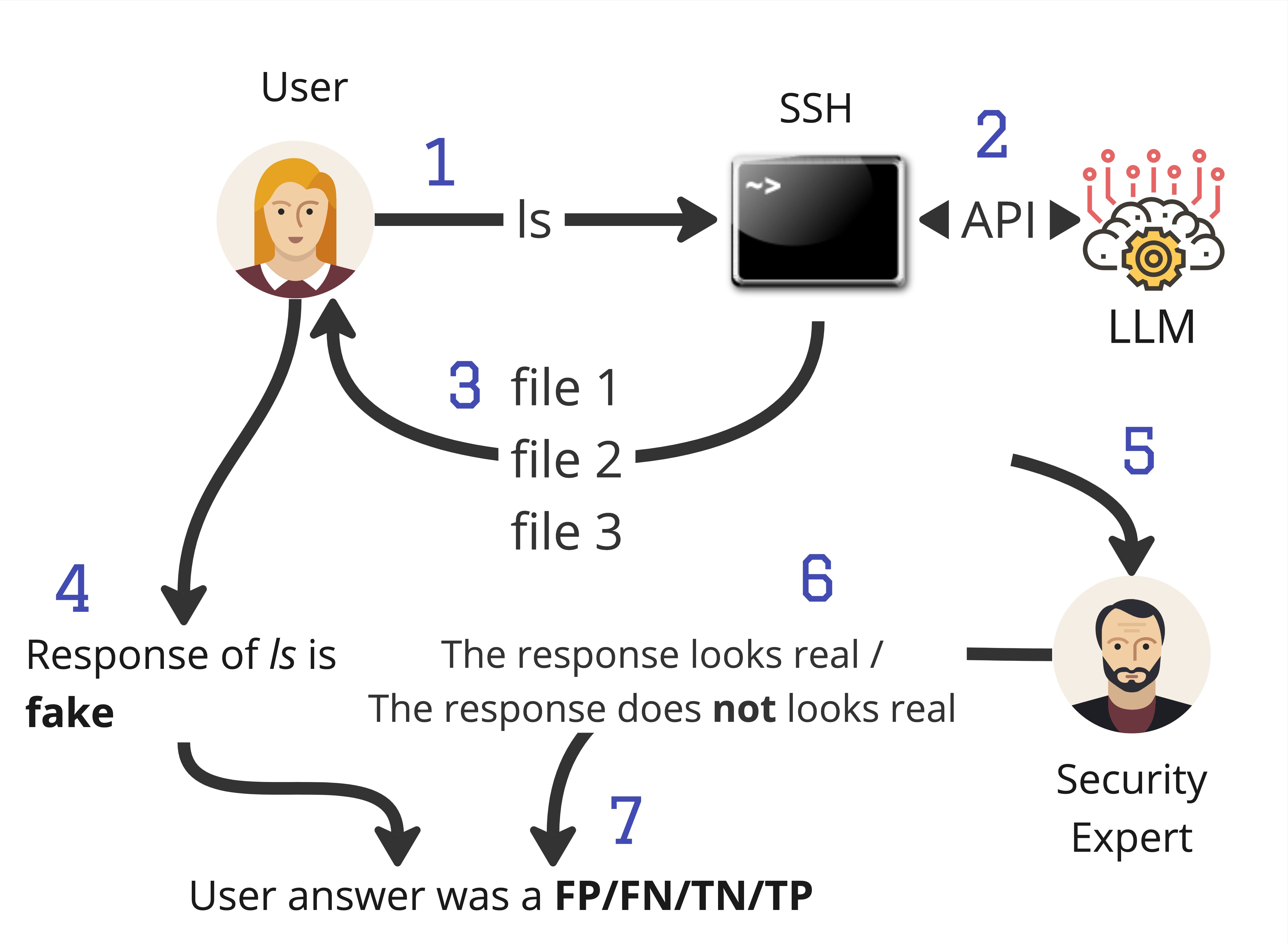}
   \caption{Evaluation of \texttt{shelLM}. Participants interact and send answers with their assessment. Security experts evaluate the outputs of \texttt{shelLM} and the answers, determining if it was a FP/FN/TN/TP.}
    \label{fig:honeypot-evaluation}
\end{figure}

Answers were manually processed to obtain a list of input commands, their outputs, and the participant's answer. To verify the results, we consulted with three Linux cybersecurity experts, who checked the correctness of the output and manually evaluated each response. Figure~\ref{fig:honeypot-evaluation} shows the evaluation diagram.

\subsection{Type Error Interpretation}
\label{subsec:error}
The error types we used have a very unique interpretation given the particular nature of our evaluation. Recall that \textbf{the goal of our experiments was not to evaluate if a participant could identify a honeypot or not; but to evaluate if the participant, knowing that it is a honeypot, could successfully identify what parts of the answers disclose the presence of a honeypot.} 

The interpretation of each error type for our case is: 
\begin{itemize}
    \item True Positive (\textbf{TP}): Participants say that the command response does not match the expected output of a Linux shell (positive for honeypot), and the expert agrees the command response is not consistent with a real Linux shell (true). 
    \item False Positive (\textbf{FP}): Participants say that the command response does not match the expected output of a Linux shell (positive for honeypot) but the expert disagrees and considers the output to be consistent with a Linux shell (false).
    \item False Negative (\textbf{FN}): Participants say that the command response does match the expected output of a Linux shell (negative for honeypot), but the expert believes that the response was not consistent with the output of a real Linux shell. This is considered beneficial for the honeypot system.
    \item True Negative (\textbf{TN}): Participants say that the command response does match the expected output of a Linux shell (negative for honeypot) and the expert agrees that it was consistent with the output of a Linux shell (true).
\end{itemize}

The following example shows a true positive, in which a participant believed the response was not consistent with the output of a Linux shell and identified the system as a "honeypot", and experts confirmed that it was a clear LLM-introduced error.

\begin{tcolorbox}
\begin{small}
\begin{verbatim}
jennifer@itcompany:~$ w
bash: syntax error near unexpected 
token `newline'
\end{verbatim}
\end{small}
\end{tcolorbox}

The example below shows a false positive, in which a participant identified the system as a "honeypot", but experts confirmed that it is common that the '.bashrc' file does not exist in systems that use other interpreters, for example, in Linux-based IoT devices. This answer may be an artifact of telling the participants they would be inside a honeypot. 

\begin{tcolorbox}
\begin{small}
\begin{verbatim}
jennifer@itcompany:~$ cat .bashrc
cat: .bashrc: No such file or directory
\end{verbatim}
\end{small}
\end{tcolorbox}

Another example of a false positive was the command shown below, in which the participant identified this command as not consistent with the output of a real Linux shell and classified it as \textbf{a honeypot} due to the small number of files. However, many small systems have this number of files or fewer. 

\begin{tcolorbox}
\begin{verbatim}
jennifer@itcompany:~$ ls /var/run
acpid.pid crond.pid 
dhclient-eth0.pid  dhclient.pid 
initramfs motd sshd.pid 
sudo utmp wpa_supplicant.pid
\end{verbatim}
\end{tcolorbox}

\subsection{Metrics for Evaluating Honeypot Software}
Following the error interpretation described in subsection~\ref{subsec:error}, the following metrics are  considered for evaluating the honeypot software performance:

\noindent \textbf{Overall Accuracy:} indicates the number of command responses correctly identified by the participants, over all the command responses received by all the participants.     

\noindent  \textbf{False Negative Rate:} It is the ratio between the number of answers that wrongly said it did match a Linux shell over all the answers that were genuinely generated matching a Linux shell. It indicates that the honeypot is effective at masquerading as a Linux shell system.

\noindent  \textbf{True Negative Rate (Specificity):} It is the ratio between the number of answers that \textbf{correctly} said it did match a Linux shell over all the answers that really matched a Linux shell. It indicates how effective the answers are to appear as a real Linux shell. 

\noindent  \textbf{False Discovery Rate:}  It is the ratio between the number of answers that \textbf{correctly} said they did not match a real Linux shell over the total number of times the users said it did not match a real Linux shell. It is an important metric to minimize because it measures the number of times the honeypot is \textit{correctly} identified as not matching a real Linux shell.

\section{Results} 
A total of 12 human participants tested \texttt{shelLM}. Participants executed 226 commands associated with package management, file system management, network operations, system information, and file management. Only 76 unique commands were executed. The top ten commands executed were: \texttt{cat}, \texttt{ls}, \texttt{sudo}, \texttt{get}, \texttt{echo}, \texttt{pwd}, \texttt{nano}, \texttt{ping}, \texttt{ssh}, and \texttt{whois}. On average, each participant executed 19 commands, with a minimum of 12 and a maximum of 38 commands.

The results of the experiments are summarized in Figure~\ref{fig:performancemetrics}. The analysis showed that 74\% (167) of the commands that seemed credible to participants had realistic outputs (True Negative Rate). A total of 7.5\% (17) of commands flagged as honeypot indicators were false positives and could be attributed to the knowledge of participants that they are in a honeypot system (False Discovery Rate). An 18\% (41) of commands flagged as honeypot indicators produced incorrect or strange output (True Positives). The remaining 0.5\% (1) of commands were false negatives, in which participants missed noticing inconsistencies in the model output.

\begin{figure}[h!]
\begin{tcolorbox}
\begin{verbatim}
            Experts
Participants REAL FORGED
REAL          167      1
FORGED         17     41
                                         
         Accuracy : 0.9204         
           95% CI : (0.877, 0.9521)
No Information Rate : 0.8142         
P-Value [Acc > NIR] : 5.432e-06      
                     
False Negative Rate  : 0.0232    
True Negative Rate   : 0.9076         
False Discovery Rate : 0.0923                                     
'Positive' Class : REAL          
                                   
\end{verbatim}
\end{tcolorbox}
\caption{Confusion Matrix and Performance metrics for the LLM-based honeypot.}
\label{fig:performancemetrics}
\end{figure}

For our purposes, a good honeypot would have True Negatives and False Negatives since we would need participants to say \textit{negative}, meaning the responses would \textit{not reveal a honeypot}. This is different from other detection methods, where what is expected is mostly a large number of true positives. If we look at the results per participant shown in Table~\ref{tab:performancemetrics}), we can confirm that the True Negative Rate was 1 in most cases. This indicates that the honeypot effectively convinces participants that the output is consistent with a Linux shell. On the other hand, \texttt{Participant 9} is a notable outlier with a TNR of $0.5$, suggesting the output was only partially successful in looking like a real Linux shell for this participant. 

The FNR values are generally low, which demonstrates the system's effectiveness. 

Finally, when considering the overall accuracy, most participants have an accuracy above $0.9$, showcasing again the general effectiveness of the honeypot.

\begin{table}[h!]
\centering
\caption{Performance metrics of the LLM honeypot per participant}
\label{tab:performancemetrics}
\begin{tabular}{@{}rrrrr@{}}
\toprule
\textbf{Participant} & \textbf{Accuracy} & \textbf{}\textbf{FNR}   & \textbf{TNR} & \textbf{FDR}    \\
\midrule
1  & 0.643 & 0.417  & 1   & 0 \\
2  & 0.941 & 0.071  & 1   & 0 \\
3  & 1     & 0      & 1   & 0 \\
4  & 0.882 & 0.2    & 1   & 0 \\
5  & 1     & 0      & 1   & 0 \\
6  & 0.923 & 0.090  & 1   & 0 \\
7  & 0.923 & 0.105  & 1   & 0 \\
8  & 0.933 & 0.090  & 1   & 0 \\
9  & 0.938 & 0      & 0.5 & 0.5 \\
10 & 0.867 & 0.167  & 1   & 0 \\
11 & 0.75  & 0.273  & 1   & 0 \\
12 & 1     & 0      & 1   & 0 \\
\bottomrule
\end{tabular}
\end{table}

Widely common commands such as \texttt{ls}, \texttt{pwd}, and \texttt{whoami} always returned the expected output. Similarly, commands including \texttt{ping}, \texttt{wget}, \texttt{whois}, \texttt{tcpdump}, \texttt{sudo -su} and non-valid command inputs resulted in credible output. The issues that occurred were inconsistencies between files in \texttt{/proc} and \texttt{/etc} directories, possibly caused by the limitation on the number of tokens and the occasional strange behavior of the LLM. This strange behavior of the model can be attributed to its context getting bigger. As it was observed recently~\cite{liu2023lost}, Large Language Models perform the best when the relevant information is at the beginning or the end of the input context. When the relevant information is in the middle of the input context, the model's performance tends to degrade.

\section{Cost analysis}
\label{sec:cost-analysis}
A primary constraint with the use of \texttt{shelLM} is the cost involved in the generation process. Since the current implementation is based on the OpenAI GPT-3.5-turbo-16k model that has high financial resource requirements, the use of this technology in every scenario should be carefully considered. The cost of generating responses using GPT-3.5-turbo-16k can add up quickly, especially when processing large volumes of text. 

At the time of the experiments, the cost for the GPT-3.5-turbo-16k model was $\$0.003$ for 1k input tokens and $\$0.004$ for 1k output tokens~\cite{OpenAIpricing}. The cost estimate for one full \texttt{shelLM} session of 16k tokens is around $\$0.70$, from which $\$0.658$ can be attributed to the input tokens and $\$0.042$ to the output tokens. 

In Table \ref{tab:experimentcost}, we present a comprehensive cost analysis for each of the 12 participants in the experiments. The table presents, for each participant, the total token count, input and output costs, and total session duration.

The average duration participants spent interacting with \texttt{shelLM} was around 35 minutes. Based on this data, we can say that the cost of usage of shelLM is around \$0.4 per 30 minutes of active use, which is roughly \$0.8 per hour. The total cost of our experiment was \$5.29, out of which \$4.99 was for input tokens and \$0.30 for output tokens. Since shelLM was designed to be mainly used inside a private network, therefore the number of expected attackers is considerably low.

The initial personality prompt was a fundamental element for simulating a Linux shell. Since it was included in the prompt of all the participants of the experiments, we analyzed its costs independently. The initial personality prompt was composed of 2138 tokens, and its cost was \$0.0065.

Based on this cost calculation, we estimate that the costs are not prohibitive. Also, in the future costs of using LLMs might decrease even further. However, researchers must be aware that an abuse of the cost of shelLM can happen if the system uses too many tokens, so we recommend a limit on the budget of the account used by shelLM.

\begin{table}[h!]
\centering
\caption{Costs for running the experiment per participant}
\label{tab:experimentcost}
\begin{tabular}{@{}rrrrr@{}}
\toprule
\textbf{Participant} & \textbf{Tokens} & \textbf{Input \$} & \textbf{Output \$} & \textbf{Time [m]} \\
\midrule
1  & 7243 & 0.2939 & 0.0189 & 47 \\
2  & 7455 & 0.3120 & 0.0197 & 27 \\
3  & 9572 & 0.4224 & 0.0252 & 24 \\
4  & 12731 & 0.5616 & 0.0355 & 26 \\
5  & 14365 & 0.6334 & 0.0379 & 44 \\
6  & 9098 & 0.4012 & 0.0241 & 31 \\
7  & 12306 & 0.5418 & 0.0324 & 42 \\
8  & 5253 & 0.2317 & 0.0138 & 34 \\
9  & 9586 & 0.4225 & 0.0253 & 55 \\
10 & 3963 & 0.1748 & 0.0104 & 12 \\
11 & 12831 & 0.5654 & 0.0338 & 53 \\
12 & 9755 & 0.4295 & 0.0257 & 29 \\
\bottomrule
\end{tabular}
\end{table}
 
\section{Conclusion and future work}
We presented \texttt{shelLM}, a novel application of LLMs for shell-based honeypots, resulting in a system that dynamically generates synthetic data as demanded by the user. The core of \texttt{shelLM} was implemented using several LLM prompt engineering techniques. Twelve security experts tested the system, and three security experts evaluated the honeypot's correspondence with a real Linux shell. During this evaluation, \texttt{shelLM} obtained a TNR of $90\%$. The results confirmed our hypothesis that it was possible to create an LLM honeypot that humans find hard to distinguish from a real system. 

Despite the encouraging preliminary results, we are aware of the limitations of our current implementation. In particular, the occasional strange behavior of the LLM is caused by its inherent stochastic nature and its memory issues related to the size of the input context~\cite{liu2023lost}. Answer latency due to the API responsiveness and the response generation time are two other limitations of the current implementation. This might be an indicator of a honeypot, however, not strong enough to make a definite conclusion, since there exist IoT devices, that have slower response times as well.

Future work will improve LLM-based honeypot responsiveness, engagement, and error management. We plan to use local LLM models and to fine-tune the models to attempt to remediate forgetfulness and behavior deterioration issues with a larger context. In addition, we intend to compare \texttt{shelLM} with other well-known honeypots to determine its appeal to attackers and the quality of data it produces. More experiments will be run where the participants would not know that they are being tested in the recognition of honeypots to measure their growing suspiciousness. Finally, we plan to conduct more experiments focused on differentiating human from bot behaviors within the honeypot and identify their respective signals. Since automated attacks are a common practice we plan to asses how effective \textit{shelLM} is in those cases and compare it with the effectiveness against human attacks. We also plan to further understand its behavior against actual attacks by deploying it in the wild.

\bibliographystyle{IEEEtran}
\bibliography{bibliography}

\end{document}